\documentstyle[aps,preprint,epsf,epsfig,rotate,prl]{revtex}
\tightenlines
\newcommand{\be}{\begin{equation}}
\newcommand{\ee}{\end{equation}}
\newcommand{\ba}{\begin{eqnarray}}
\newcommand{\ea}{\end{eqnarray}}

\draft        
\begin{document}             
\title{Optical absorption in a degenerate Bose gas}
\author{S. K. Yip}
\address{Institute of Physics, Academia Sinica, Nankang, Taipei 11529, Taiwan}
\maketitle          

\begin{abstract}

We here develop a theory on optical absorption in a dilute Bose gas
at low temperatures.  This theory is motivated by the Bogoliubov
theory of elementary excitations for this system, and takes 
into account explicitly the modification of the nature and
dispersion of elementary excitations due to 
Bose-Einstein condensation. 
Our results show important differences from existing theories.

PACS numbers: 03.75.Fi, 32.70.Jz, 32.80.-t
\end{abstract}

\vskip 0.2 cm

A remarkable property of a degenerate Bose gas is the 
modification of the nature of excitations in the system.
This has been elucidated by the work of Feymann and Bogoliubov
\cite{B47}.
For a classical or non-degenerate Bose gas, elementary excitations
are basically quasiparticles.  For a degenerate Bose gas however,
the long wavelength elementary excitations are the same as
density waves of the system, propagating at the sound
velocity.  They correspond to neither addition nor removal
of a quasiparticle, but a linear combination of both.  At
short wavelengths these excitations resemble more closely
 quasiparticles, with the dispersion the same as
 free particles apart from a shift. 

Recently there is much interest in optical excitations 
in dilute degenerate Bose gas.  The frequency of 
absorption has been used
as an indication of the formation of Bose-Einstein condensation
in the system \cite{Fried98}.  An optical excitation
is a non-trivial process, as one of the original atoms
(referred to as a-atoms below), 
all identical before the excitation occurs, is 
converted to another one which is distinguishable from all
the others.  Moreover a-atoms which are not excited internally 
 interact differently with
this `foreign' atom (referred to as c-atom below) than 
among themselves, and must
respond by rearrangement of their relative motion.
The frequencies at which optical absorption occur
are therefore different from $\omega_0$, the value 
for a single isolated atom.

Optical absorption in a Bose gas has been considered by Oktel and
Levitov \cite{Oktel99},
and also by Pethick and Stoof \cite{Pethick01}.
Johnsen and Kavoulakis \cite{Johnsen01} considered
the case of excitons in semiconductors where
the initial and final bosons have different effective masses, using
an approximation equivalent to that used by Oktel and Levitov \cite{Oktel99}. 
Let us briefly summarize the main results of Ref.
\cite{Oktel99}  most relevant to the discussions
below.  These authors concentrate on intermediate to
high temperatures and ignore the modifications of the
nature of quasiparticles below the transition temperature.
They make the dramatic prediction that, for $T < T_c$
and in the limit of no momentum transfer,
there is actually not one but {\it two} absorption lines.
These lines are located at the frequencies
\be
\omega = \omega_0 +  g_{ac} n - g_{aa} (n_0 + 2 n_T)  + y
\label{ol}
\ee
where $y$ obeys the quadratic equation 
$y ( y - g_{ac} n_T + g_{aa} n_0) = g_{ac}^2 n_0 n_T $.
Here $ g_{aa}$ and $g_{ac} $ are the interaction parameters 
among the a-atoms and between c- and a-atoms respectively,
$n_0$ the number density of the condensate and $n_T$ that of the 
thermal atoms, $ n = n_0 + n_T$ the total number density.
As $T \to 0$, $y \to 0$ and $ - g_{aa} n$ though
the weight of the second line also approaches zero.
For $T \to T_c$,  $ y \to 0$ and $y \to g_{ac} n$ with
the weight of the first line vanishing.  
Note then
the second solution of $y$ then yields the line
$\omega = \omega_0 + 2 (g_{ac} - g_{aa} ) n $.

Here we would like to re-examine and extend this investigation to
low temperatures.  In particular, we are interested
in the effects brought about by the
macroscopic occupation of the lowest energy state.
Note that not only the dispersion is changed for $T < T_c$,
but also that now the annihilation of an a-atom can
correspond both to an absorption and a creation 
of a Bogoliubov excitation.  
We shall see that properly accounting for these features
substantially modifies the absorption spectrum. In general 
it no longer simply consists of two lines.  
In some regimes where the spectrum {\it is} approximately two 
relatively sharp lines,
the frequencies of these lines differ from
those of eq (\ref{ol}).

We then consider a weakly interacting Bose gas.  The Hamiltonian
of this system is given by
\be
H_a = \sum_p \xi_p  a^{\dagger}_{p} a_{p}
  + {g_{aa} \over 2V} \sum' a^{\dagger}_{p1} a^{\dagger}_{p2} a_{p3} a_{p4}
\label{Ha}
\ee
Here $\xi_p = p^2 / 2m$ is the kinetic energy of a free
a-boson of momentum $p$
(we shall drop all vector labels since no confusion will arise),
and the second sum is over $p$'s with the constraint
$p_1 + p_2 = p_3 + p_4$.  $V$ is the volume.
The Hamiltonian of the system containing
also a c-atom is given by
$H = H_a + H_c + H_{ac}$.
$H_c$ 
describes the free motion of the c-atom
\be
H_c = \sum_p ( \xi_p + \omega_0) c^{\dagger}_{p} c_{p}
\label{Hc} 
\ee
where we have neglected the effective mass change in
the optical transition.  $\omega_0$ accounts for
the internal energy difference between the a and c atoms.
$H_{ac}$  describes 
the interaction of the c-atom with the rest of the a-atoms.
We write it as
\be
H_{ac} =  {g_{ac} \over V} \sum_{k',q,p'} c^{\dagger}_{k'-q} c_{k'}
           a^{\dagger}_{p'} a_{p'-q}
\label{Hac}
\ee

The absorption with momentum transfer $k$
(from the external optical perturbation)
at frequency $\omega$ is proportional to
 $ - 1 / \pi$ times the
imaginary part of the propagator $\sc D (\omega)$ whose
expression in real time is given by
\be
{\sc D} (t) = - i < [ X (t) ,  X^{\dagger} (0) ] > \theta (t)
\label{D}
\ee
where $X \equiv c_k a^{\dagger}_0 +  \sum_{p \ne 0}
               c_{k+p} a^{\dagger}_{p}$. $\theta(t)  = 1$ if
$t > 0$ and vanishes otherwise.
For here and below the angular bracket denotes the equilibrium 
expectation value at temperature $T$ 
for a system with no c-atoms under the Hamiltonian $H$.

As in \cite{Oktel99} we are led to consider the equation of
motion of the operator $X(t)$ and hence $c_k a^{\dagger}_0$ and
$c_{k+p} a^{\dagger}_p$.  To motivate the procedure
below let us consider the operator $c_{k+p} a^{\dagger}_p$
 and the contribution
to its equation of motion from
its commutator with $H_a$ 
describing the interaction among the a-atoms.  The operator
$c_{k+p}$ behaves simply as a scalar, and the terms that are
produced, apart from this factor of $c_{k+p}$, 
are identical with that for  the operator $a^{\dagger}_p$
 in a BEC of the a-atoms alone.
Thus we have a linear combination of terms involving,
among others, 
(with $c_{k+p}$ implicit)
 $a^{\dagger}_0 a_0 a^{\dagger}_{p}$,
 $a^{\dagger}_0 a^{\dagger}_0 a_{-p}$, and
$a^{\dagger}_{p'} a_{p'} a^{\dagger}_{p}$ ($p' \ne p$).
Explicitly, we have
\be
i { \partial \over \partial t} ( c_{k} a^{\dagger}_0 ) 
 =  - c_k  \left( 
{g_{aa} \over V}  a^{\dagger}_0 a^{\dagger}_0 a_0 
   + 2 { g_{aa} \over V} \sum_{p' \ne 0} a^{\dagger}_{p'} a_{p'}
   a^{\dagger}_0  \right)  + \ ...
\ee
\be
i { \partial \over \partial t} ( c_{k+p} a^{\dagger}_{p} ) 
 =  - c_{k+p}  \left( \xi_p a^{\dagger}_p   + 
 2 {g_{aa} \over V}  a^{\dagger}_p ( a^{\dagger}_0 a_0 
   +  \sum_{p' \ne 0} a^{\dagger}_{p'}  a_{p'} )  
 +  {g_{aa} \over V}  a_{-p} ( a^{\dagger}_0 a^{\dagger}_0) 
   \right)  + \ ...
\ee
where the ellipses represent contributions from the other parts
of the Hamiltonian.  
We shall treat the mentioned terms by exactly the same approximation
as in Popov's generalization of Bogoliubov's theory 
to finite temperatures \cite{Popov},
{\it i.e.}, $a_0$ and $a^{\dagger}_0$
are replaced by scalars  ( $ \sqrt{N_0} e^{ \mp i \mu t } $ ) and 
$a^{\dagger}_{p'} a_{p'} a^{\dagger}_{p}$
is replaced by $ < a^{\dagger}_{p'} a_{p'} > a^{\dagger}_{p}$
where the angular bracket represents equilibrium expectation values.
Note that thus accordingly the chemical potential
 $ \mu =  g_{aa} ( n_0 +  2 \sum_{p \ne 0} n_p ) $
where $N_p = < a^{\dagger}_{p} a_{p} > $ is the number  
of particles at momentum $p$ and $n_p$ the corresponding density.
We are thus led to the conclusion that 
the operator $c_{k+p} a_{-p}$ is generated automatically.
This operator was not taken into account in ref \cite{Oktel99}.
 Also, if the Hamiltonian were to consist of $H_a$
alone, these equations, together with those
with $a_p^{\dagger}$ replaced by $a_{-p}$,   can be diagonalized by the
Bogoliubov transformation with the excitation operators a linear
combination of $a_{p}$ and $ a^{\dagger}_{-p}$
(with coefficients identical to those of a system of a-atoms only).

The commutators of $c_k a^{\dagger}_0$ and
$c_{k+p} a^{\dagger}_p$ with $H_c$ are simple since $a$'s and
$a^{\dagger}$'s then act like scalars.  These
terms simply describe the free motion of the c-atom.
Finally, we consider in some detail the contributions
from $H_{ac}$ and emphasize another crucial difference
of our present treatment from that of \cite{Oktel99}.
Consider  $c_k a^{\dagger}_0 H_{ac}$.  The contributions
from the $q=0$ term in eq (\ref{Hac}) simply gives rise to
$g_{ac} n$ times the original operator.  For 
$q \ne 0$, we have ${g_{ac} \over V}
 \sum_{k'} c_k c^{\dagger}_{k'-q} c_{k'} 
   a^{\dagger}_0 ( a^{\dagger}_q a_{0} + a^{\dagger}_0 a_{-q} 
        + \sum_{p' \ne 0, q} a^{\dagger}_{p'} a_{p'-q} ) $
where we have split off explicitly terms involving the
condensates.  Since there is at most one c-atom
this expression has a finite contribution only when $k'-q = k$.  
Thus we have terms of the form
$ c_{k+q} a^{\dagger}_0  ( a^{\dagger}_q a_{0} + a^{\dagger}_0 a_{-q} )$
involving explicitly the condensate.   This has the obvious
interpretation (see more details below):
  the c-atom can be scattered from $k+q$ to $k$
in two ways in the presence of a condensate,
either by removing/adding one condensate atom
while creating/annihilating another at $+q$/$-q$.
Note that each of these in turn can correspond to 
emission/annihilation of an excitation at $+q$/$-q$.
  Again replacing $a_0$ and $a_0^{\dagger}$ by
scalars implies that one needs to take into account
the operator $c_{k+p} a_{-p}$.

The above leads us to evaluate the equations of motion
of the propagators
$ - i \sqrt{N_0} < [c_k (t),  X(0) ] > \theta(t) $,
$ - i < [c_{k+p} a^{\dagger}_p (t),  X(0) ] > \theta(t) $, and
$ - i < [c_{k+p} a_{-p} (t),  X(0) ] > \theta(t) $
($ p \ne 0$) .
We shall do so under the generalization of the Bogoliubov-Popov
approximation as explained earlier.
Alternatively, we can introduce from the start the
new operators $\alpha_p$ by $ a_p e^{i \mu t } = u_p \alpha_p - v_p \alpha_{-p}^{\dagger}$, {\it the same} Bogoliubov transformation
 as in the corresponding pure a-atom system
 ( For simplicity, we shall choose the gauge where
$u_p$ and $v_p$ are all real.  Note also $u_p$ and $v_p$  
are both even in $p$.)
We then consider the three propagators
$ \eta_k^{(0)} \equiv - i \sqrt{N_0} < [c_k (t),  X(0) ] > \theta(t) $,
$ \eta_{p,k}^{(1)} \equiv i < [c_{k+p}  {\alpha}^{\dagger}_p (t),  X(0) ] > 
\theta(t) $,
 and
$ \eta_{p,k}^{(2)} \equiv - i < [c_{k+p} {\alpha}_{-p}(t),   X(0) ] > 
\theta(t) \  $.
This has the advantage that the part due to $H_a$ is already
diagonalized. 
These propagators are related to the probability amplitude that, at time $t$,
the system is in the state with an exciton consisting of, respectively,
annihilation of a condensate atom at $p=0$ with the c-atom at $k$,
annihilation of an excitation of momentum $p$ with the c-atom at $k+p$,
and creation of an excitation of momentum $-p$ with the c-atom also
at $k+p$.  We shall call these `excitons' 
type 0, 1 and 2 respectively.
The resulting equations of motion read, with ${\omega}' 
\equiv \omega - ( \omega_0 + g_{ac} n - \mu)$.

\be
({\omega}' - \xi_k ) \eta_k^{(0)}
= g_{ac} {N_0  \over V} 
          ( \eta_k^{(\mu)} - \eta_k^{(\nu)} ) + N_0
\label{e0}
\ee

\be
({\omega}' - (\xi_{p+k} - E_p) ) \eta_{p,k}^{(1)}
= g_{ac} {\tilde N_p  \over V}  ( u_p - v_p ) \eta_k^{(0)} \ + \ 
       g_{ac} { \tilde N_p \over V}   
    ( u_p \eta_k^{(\mu)} +  v_p \eta_k^{(\nu)} ) 
 \  + \ u_p \tilde N_p
\label{e1}
\ee

\be
({\omega}' - (\xi_{p+k} + E_p) ) \eta_{p,k}^{(2)}
= g_{ac} {\tilde N_p + 1  \over V}  ( u_p - v_p ) \eta_k^{(0)} \ - \ 
       g_{ac} { \tilde N_p + 1 \over V}   
    ( v_p \eta_k^{(\mu)} +  u_p \eta_k^{(\nu)} ) 
 \  - \ v_p ( \tilde N_p + 1 )
\label{e2}
\ee

\noindent where $\tilde N_p = < \alpha_p^{\dagger} \alpha_p > =
 { 1 \over e^{E_p/T} - 1 }$
is the bosonic distribution function
(the number of excitations,  {\it not particles}),
$E_p = \sqrt { \xi_p^2 +  2\xi_p ( g_{aa} n_0) }$
the excitation energies under the Bogoliubov-Popov approximation,
and we have introduced the short hands
$ \eta_k^{(\mu)} = \sum_{p \ne 0} 
 ( u_p \eta_{p,k}^{(1)} -  v_p \eta_{p,k}^{(2)} ) $
and
$ \eta_k^{(\nu)} = \sum_{p \ne 0} 
 ( v_p \eta_{p,k}^{(1)} -  u_p \eta_{p,k}^{(2)} ) $.
Note that $\omega'$ is identical with $y$ of eq (\ref{ol}) if one
replaces $n_T$ by $\sum_p n_p$ and
uses the mentioned approximation for the chemical potential.

The interpretation of these equations is clear by an examination
of their form.  In the absence of $H_{ac}$, the excitons
0, 1, 2 have energies (measured with respect to
$\omega_0 + \mu$) $\xi_k$, $\xi_{p+k} - E_p$, and $\xi_{p+k} + E_{p}$
respectively.  These different possibilities of excitons
 are coupled together by the interaction $g_{ac}$.
The factors $\tilde N_p$ and $\tilde N_p + 1$ are the bosonic factors
associated the absorption or
emission of an excitation at $ \pm p$.  Note that
even in the absence of the Bogoliubov transformation
(putting $u_p = 1$ and $v_p = 0$ by hand), $\eta^{(2)}_{p,k}$ is 
still finite:  it can be generated from a type 0 exciton by
emission of an excitation to $-p$, and is allowed so long
as $N_0$ is macroscopic (note that the definition of
$\eta^{(0)}_k$ contains the factor $\sqrt N_0$. )
These type 2 excitons can in turn be converted back to
type 0, as explicitly shown in eq (\ref{e0}).
These processes were not included in the calculations of
ref \cite{Oktel99} and \cite{Johnsen01}.

Let us first check how our results reduce to those of \cite{Oktel99}
for $ T > T_c$.  In this regime there is no need to 
treat $\eta_k^{(0)}$ in a special manner 
since $N_0$ is no longer
macroscopic.  Also $u_p = 1$ and $v_p = 0$.  In this case
we need only $\eta^{(1)}_{p,k}$ and equation (\ref{e1}) becomes
equivalent to that used in Ref. 
 \cite{Oktel99}. For $k=0$  there is only one line and
the frequency is given by,
from eq (\ref{e1}),
$\omega' = g_{ac} n $.  Since $\mu = 2 g_{aa} n$,
$\omega =  2 ( g_{ac} - g_{aa}) n$, reproducing
the results of \cite{Oktel99} and \cite{Pethick01}.

For $T < T_c$  we need to consider 
the three coupled linear equations
(\ref{e0}-\ref{e2}).  The terms on the right hand side
involving explicitly $g_{ac}$ represent `vertex corrections'
if the above calculations were formulated in terms of
Feymann diagrams.  They in general cannot be ignored.
However, it is of interest to pretend that this can be
done and examine the results.  This is a good approximation
at very low temperatures for small gaseous parameters ( see Appendix).
  For simplicity
we shall discuss only $k=0$ below. In this case $\eta_k^{(0)}$
has a pole at $\omega' = 0$.  $\eta_{p,k}^{(1)}$ however
has a pole whose location depends on $p$.  It is responsible for transitions
with $\omega' $ ranging from $0$ for $p \to 0$ to 
$ - g_{aa} n_0$ for large $p$.
 ( this is due to  the shift between $\xi_p$ and $E_p$ 
as $p \to \infty$).    If we ignore the modification of
the spectrum at small $p$'s 
we have in total two lines, one at $\omega' = 0$ and
the other at $ - g_{aa} n_0$, with the weight of the 
second line vanishing if $T \to 0$. This result
is similar to that of ref \cite{Oktel99} at the same low 
temperatures limit.

In general we can obtain from eq (\ref{e0}-\ref{e2}) three coupled equations 
involving $\eta_k^{(0)}$,  $\eta_k^{(\mu)}$ and $\eta_k^{(\nu)}$.
The formulas are lengthy and we shall not show them here.
The desired propagator is given by ${\sc D} = \eta_k^{(0)} + \eta_k^{(\mu)}$.

We shall show the numerical results below.
For technical simplicity we shall fix $n_0$ 
and the (modified) gaseous parameter $ (n_0 a^3)^{1/2}$
 while vary the temperature.
[ We do this to avoid solving self-consistent equations
for the chemical potential].  
In so doing the total number of particles $n$ is {\it not} fixed
when the reduced temperature $t \equiv T / g_{aa} n_0 $ varies 
( being a function of only $t$ and 
 $ (n_0 a^3)^{1/2}$).
However,  we checked that 
in the (low) temperature range investigated below the change 
in the total particle density amounts
to less than a few percent. Thus our results can still
illustrate the semi-quantitative behavior for fixed $n$.

\begin{figure}[h]
\epsfig{figure=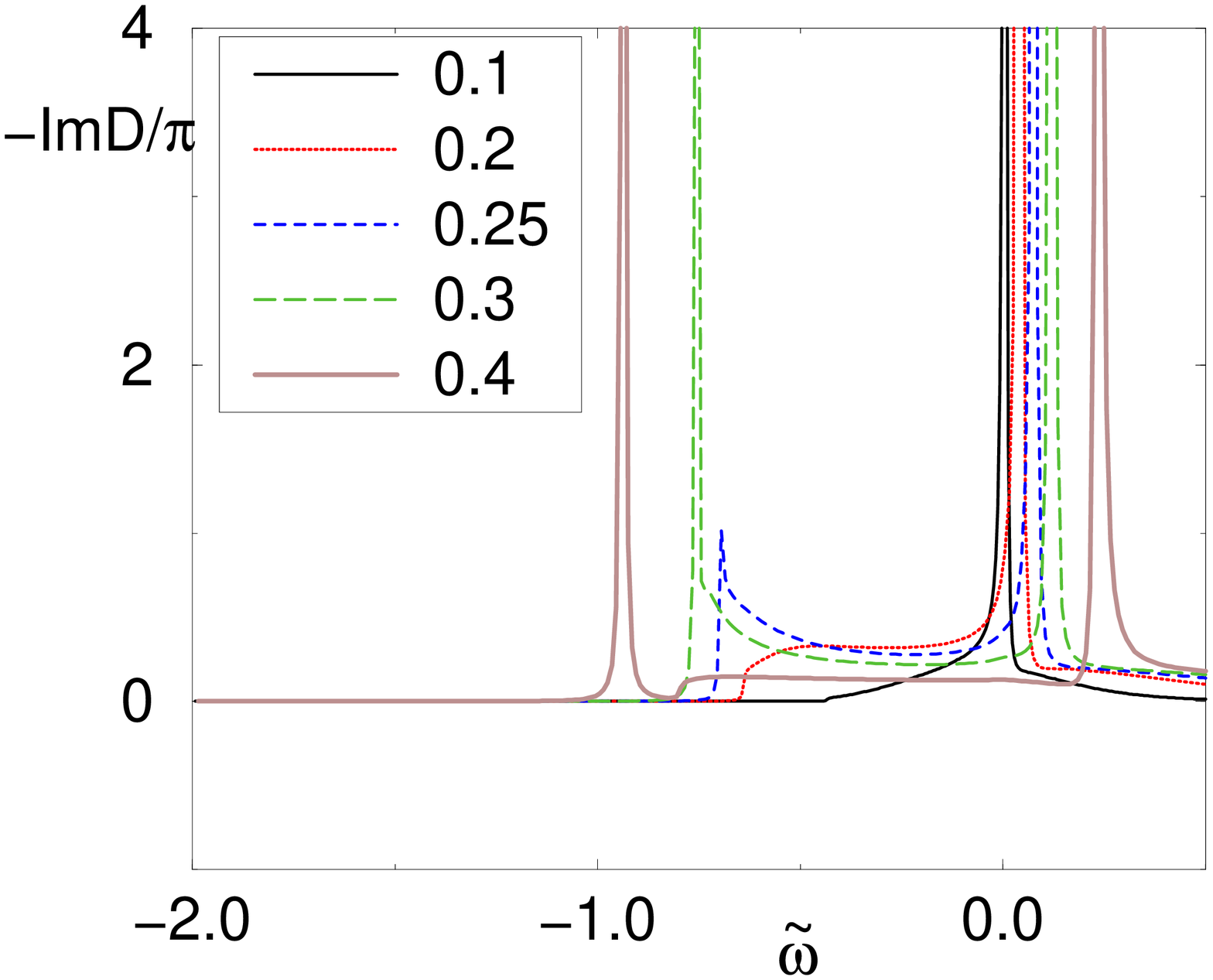,width=3in,angle=-90}
\vskip 0.2 cm
\begin{minipage}{0.45\textwidth}
\caption[]{ Absorption spectrum for $g_{ac}/g_{aa} = 20$,
$( n_0 a^3)^{1/2} = 0.01$, for reduced temperatures
$t$ shown in the legend. }

\label{fig:g20}
\end{minipage}
\end{figure}

An example for $g_{ac}/g_{aa} = 20$ is
as shown in Fig. \ref{fig:g20}.    At very low temperatures
$ t= 0.1$ the absorption is basically a sharp line at
dimensionless frequency $\tilde \omega \equiv
\omega'/ (g_{aa} n_0 ) = 0$ 
except for small wings from absorption (emission)
of quasiparticles at frequencies above (below) the main 
absorption line.  At increasing temperatures the lower frequency
wing grows, eventually evolves into something which resembles
a sharp line at a frequency which decreases with temperature.
At the same time the upper line increases in frequency.
We checked that at higher temperatures than the ones shown,
 the weight of the lower (upper) line decreases (increases).
While our calculation cannot be simply extended into
higher temperatures (see below), we believe that the upper line will
eventually evolve to the line $\omega = 2 (g_{ac} - g_{aa}) n$
for $ T \ge T_c$.

\begin{figure}[h]
\epsfig{figure=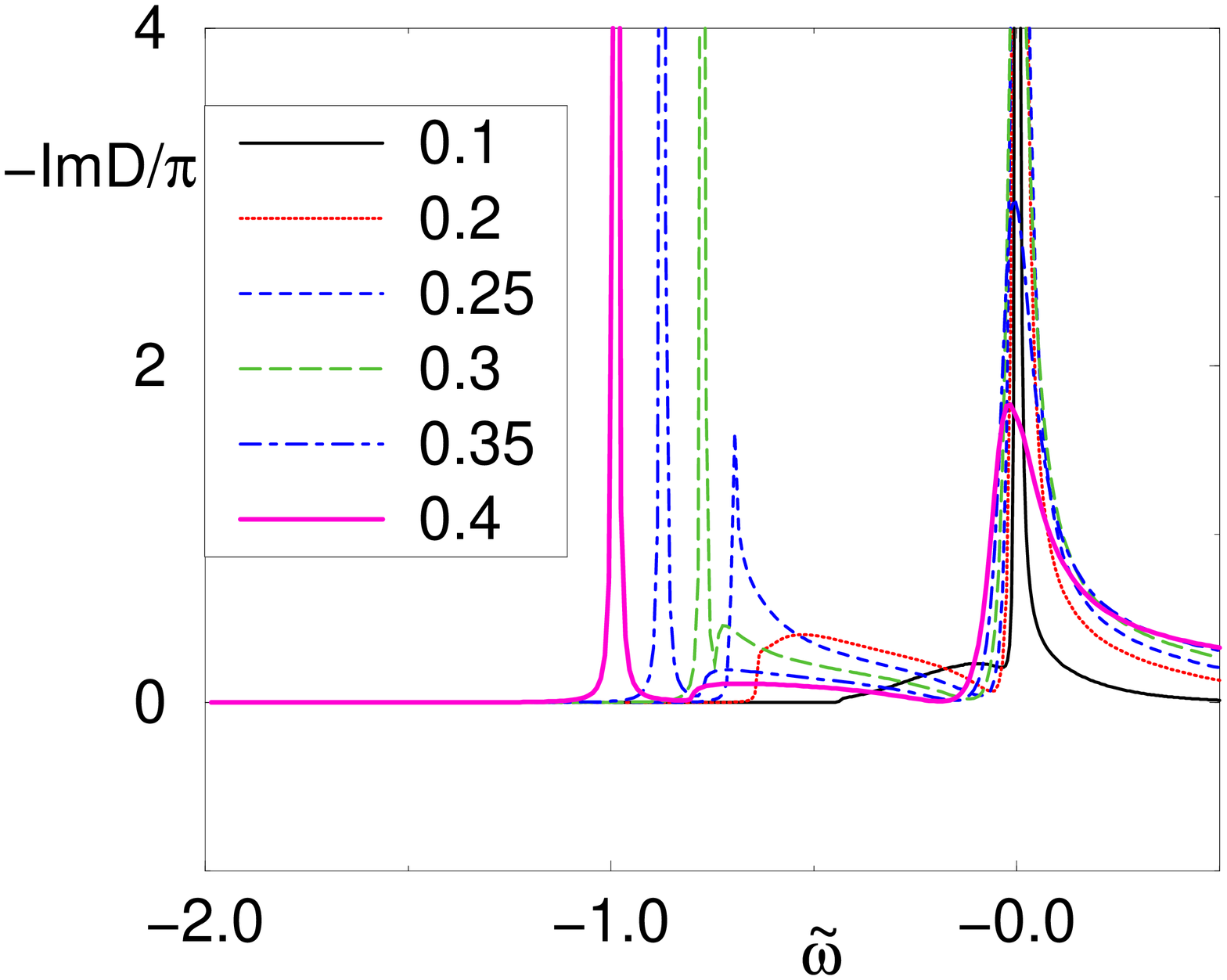,width=3in,angle=-90}
\vskip 0.2 cm
\begin{minipage}{0.45\textwidth}
\caption[]{Absorption spectrum for $g_{ac}/g_{aa} = -20$,
$( n_0 a^3)^{1/2} = 0.01$, for reduced temperatures
$t$ shown in the legend. }

\label{fig:gn20}
\end{minipage}
\end{figure}

\begin{figure}[h]
\epsfig{figure=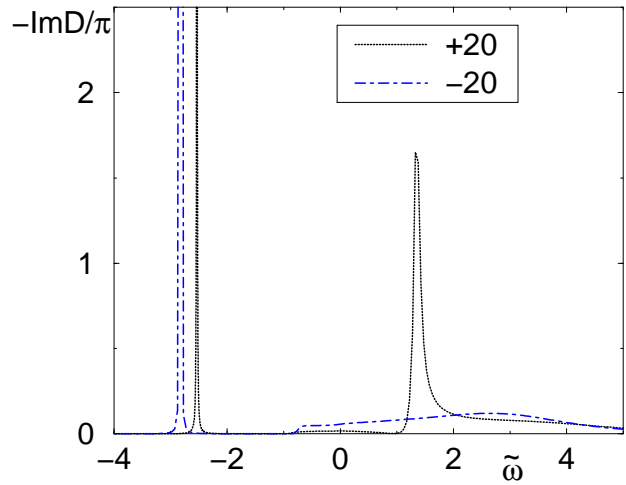,width=3in,angle=-90}
\vskip 0.2 cm
\begin{minipage}{0.45\textwidth}
\caption[]{  Absorption spectrum for $g_{ac}/g_{aa} = \pm 20$,
$t = 1.0$, for $( n_0 a^3)^{1/2} = 0.01$.  }
\label{fig:t1}
\end{minipage}
\end{figure}

The results for a large and negative $g_{ac}$ is as shown
in Fig \ref{fig:gn20}. At very low temperatures the results
are qualitatively similar to positive $g_{ac}$. However,
at temperatures as low as $t = 0.4$ the results are qualitatively
different from those of $g_{ac} >0$.  The upper line is much
broader and extends to much higher frequencies.
The dependence on the sign of $g_{ac}$ becomes even more
significant at higher temperatures, as shown in 
Fig \ref{fig:t1} for $t = 1.0$.
Note that for the 1s $\to$ 2s transition in atomic hydrogen
\cite{Fried98}, $g_{ac}/g_{aa}$ is likely to be large and negative.  
The implication of this on experiments in traps
where the density is non-uniform needs however further investigations.

Though we found basically two lines at intermediate
temperatures, we note here that the separation between
the lines are quantitatively different from
those predicted in \cite{Oktel99}.  Since the distinctions
between quasiparticles and real particles are ignored
in \cite{Oktel99} there is an ambiguity in generalizing
eq. (\ref{ol}) quantitatively, according to whether one
uses the total number of excitations, or the total
number of $p \ne 0$ particles $n'(T)$, or $n'(T) - n'(0)$
for $n_T$.  For the first choice we find that, e.g., 
for $g_{ac} / g_{aa} = 20$ at $ t = 1.0$,
$\tilde \omega $ should be given by $2.49$ and $-3.1$.
The third choice yields $3.39$ and $-3.76$, while the
second choice produces an even bigger separation
between the two lines. 
 In all cases the separations between the lines are overestimated
(c.f. Fig \ref{fig:t1}).
 The disagreement for all choices further worsens at higher temperatures.  
We believe that these
differences result from the neglect in \cite{Oktel99}
of the third channel ( `type 2 excitons').

We have also investigated the dependence of the absorption spectrum
on the (modified) gaseous parameter $(n_0 a^3) ^{1/2}$.
A smaller $ (n_0 a^3)^{1/2}$ reduces (enhances) the effect
due to the thermal excitations and the result
resembles those of a lower (higher) temperature.

The treatment here is restricted to dilute gases
at relatively low temperatures. At higher temperatures
one need to worry about more complicated
objects involving annihilation and
creation of multiple Bogoliubov excitations, and the interaction
among these various channels. 
The theory for optical absorption will become much more
complicated just as a theory of the elementary 
excitations will be at these temperatures.

This research was supported by the National Science
Council of Taiwan under grant number 89-2112-M-001-105.
This project was motivated during a stay of the author
at the Aspen Center for Theoretical Physics.  The author
would like to thank the Center for its support.

{\it Appendix} -- When solving for $\eta_k^{(\nu)}$ using eq (\ref{e2}),
there arises a divergence $ \sim  { g_{ac} \over V} 
\sum_p { 1 \over \omega' - ( \xi_{p+k} + E_p )}$ from large $p$'s. 
 This divergence has
the same origin as when one studies the scattering of two particles
interacting via a $\delta$-function potential, and can be 
cured by eliminating $g_{ac}$ in favor of the scattering length. 
After this divergence is cured, it can be shown that as $T \to 0$
the vertex corrections are of order $(n_0 a^3 )^{1/2}$ small.

\end{document}